\documentclass[aps,a4paper,twocolumn]{revtex4}

\usepackage{amsmath}
\usepackage{graphicx}
\usepackage{subfigure}
\usepackage[usenames,dvipsnames]{color}
\definecolor{darkblue}{RGB}{0,0,196}
\usepackage[colorlinks=true,linkcolor=darkblue,citecolor=darkblue,urlcolor=darkblue]{hyperref}
\usepackage{setspace}
\usepackage{multirow}
\usepackage{hyperref}
\usepackage{xcolor}
\hypersetup{
  colorlinks   = true, 
  urlcolor     = blue, 
  linkcolor    = blue, 
  citecolor   = blue 
}
\DeclareUnicodeCharacter{2212}{-}
\usepackage{graphicx}
\usepackage{amsmath,bbm}
\usepackage{amssymb,bm}
\usepackage{yfonts}
\usepackage{comment}
\usepackage[normalem]{ulem}
\begin{document}

\title{Charm-hadron reconstruction through three body decay in hadronic collisions using Machine Learning}

\author{Neelkamal Mallick$^{1}$}\email[Corresponding author: ]{Neelkamal.Mallick@cern.ch}
\author{Hadi Hassan$^{1}$}\email[]{Hadi.Hassan@cern.ch}
\author{~D.J.~Kim$^{1,2}$}\email[]{dong.jo.kim@jyu.fi}%
\affiliation{$^{1}$University of Jyväskylä, Department of Physics, P.O. Box 35, FI-40014 University of Jyväskylä, Finland}
\affiliation{$^{2}$Helsinki Institute of Physics, P.O.Box 64, FI-00014 University of
Helsinki, Finland}

\date{\today}
\begin{abstract}

Studies of heavy-quark (charm and beauty) production in hadronic and nuclear collisions provide excellent testing grounds for the theory of strong interaction, quantum chromodynamics. Heavy-quarks are produced predominantly in the initial hard partonic interactions, allowing them to witness the entire evolution process. The charm hadrons are produced in two ways. Firstly, the prompt charm hadrons which are formed from the charm quark hadronization which are produced directly from the initial hard-scatterings or the decay of other excited charm states. Secondly, the nonprompt charm hadrons which are produced from the decay of beauty hadrons. The produced charm hadrons then usually decay to light-flavor hadrons or leptons via two or three body decay. The reconstruction of charm hadrons is challenging due to the large combinatorial background as well as the difficulty of distinguishing between prompt and non-prompt charm hadrons. In this work, we use machine learning models--XGboost and Deep Neural Network--to reconstruct $\Lambda_c^{+} (udc)$ hadrons via its three body final state decay channel, $\Lambda_c^{+} \rightarrow pK^0_s$ and $K^0_s \rightarrow \pi^{+}\pi^{-}$. Using several experimentally available features of the decay daughters, these models can separate signal from background and identify prompt and nonprompt candidates with nearly 99\% accuracy. This method performs an unbinned track-level reconstruction since the $\Lambda_c$ candidates are tagged directly from their decay daughters. The necessary data for this study are simulated in pp collisions at $\sqrt{s}=13.6$~TeV using PYTHIA8 (Monash) model. 

\end{abstract}

\maketitle 

\section{Introduction}
\label{sec:intro}

Two of the world's most powerful particle accelerators, the Large Hadron Collider (LHC) at CERN, Switzerland, and the Relativistic Heavy-ion Collider (RHIC) at BNL, USA have been studying the fundamental processes of ``hadronization", which explains the formation of hadrons out of quarks, in the collisions of hadronic and nuclear matter at relativistic energies. During such collisions, for a very short duration of time ($\simeq10^{-23}$~s), quarks and gluons become the relevant degrees of freedom. The measurement of heavy-flavor (mainly charm and beauty) production, is crucial since they are produced predominantly in the initial hard partonic interactions via perturbative-quantum chromodynamics (pQCD) processes, and thus, provide excellent testing grounds for the theory of strong interaction. The QCD factorization theory predicts the transverse momentum ($p_{\rm T}$) differential production cross sections of heavy-flavor as a convolution of three terms: (i) the parton distribution functions (PDFs), (ii) the parton hard-scattering cross sections, and (iii) the heavy-flavor fragmentation functions~\cite{Collins:1989gx}. Theoretically, the heavy-flavor production cross sections can be obtained from pQCD calculations using General-Mass Variable-Flavor-Number Scheme (GM-VFNS)~\cite{Kniehl:2005mk, Kniehl:2012ti}, Fixed-Order-Next-to-Leading-Log (FONLL) approach~\cite{Cacciari:1998it, Cacciari:2012ny}, and the $k_{\rm T}$ factorization framework~\cite{Maciula:2013wg}. These models generally describe the $D$- and $B$-meson production cross sections within uncertainties over a wide range of transverse momentum and the ratios of strange and non-strange $D$ mesons~\cite{Andronic:2015wma, ALICE:2019nxm, CMS:2016plw, LHCb:2017vec}. However, the predictions for beauty productions is much closer to data than that of the charm~\cite{Andronic:2015wma}. Another important aspect of the pQCD calculations is the fragmentation function, which models the hadronization process, and parametrizes the fraction of quark energy transferred to the produced hadron. Additionally, the fragmentation fractions describe the probability of a heavy-quark having hadronized into a particular hadron species. This is based on the assumption of ``universality", meaning the fragmentation processes are independent of the collision energy and colliding species~\cite{Lisovyi:2015uqa}. The measurements of relative production of baryons and mesons (``baryon-to-meson ratio") serve as an important observable sensitive to the fragmentation process since the ratio cancels out the PDFs and partonic interactions common to heavy quarks. Specifically, the ratio of production cross sections of $\Lambda_c^{+}$ relative to $D$ mesons and $\Lambda_b^{0}$ relative to $B$ mesons can provide insight into the hadronization of fragmentation of charm and beauty quarks into hadrons, respectively. Indeed, measurements of $\Lambda_c^{+}/D^0$ ratio at the LHC~\cite{ALICE:2017thy, ALICE:2020wla, ALICE:2020wfu, CMS:2019uws, LHCb:2013xam} are found to be substantially higher than that of lower energy $e^+e^-$ collisions~\cite{ARGUS:1988hly, CLEO:1990unu, ARGUS:1991vjh, Gladilin:2014tba}. On the other hand, the ratio of $\Lambda_b^{0}$ to $B$-mesons shows a transverse momentum dependence~\cite{LHCb:2011leg, LHCb:2019fns,  HFLAV:2016hnz, ParticleDataGroup:2016lqr}. In addition, Monte Carlo event generators such as PYTHIA which includes the formation of strings as the fragmentation process with fragmentation parameters tuned to previous $e^+e^-$ and $e^-p$ collisions, significantly underestimate the $\Lambda_c^{+}/D^0$ ratio~\cite{Skands:2014pea, Christiansen:2015yqa, Bierlich:2015rha}. All of these studies challenge the ``universality" of charm and beauty fragmentation fractions at the LHC energies. 

Further, the production of charm hadrons happens in two ways, and based on the origin of constituent charm quarks, they can be either prompt or nonprompt types. Firstly, the prompt charm hadrons are formed from the hadronization of charm quarks produced directly from the initial hard-scatterings or via the decay of other excited charm states. Secondly, the nonprompt charm hadrons are produced from the decay of beauty hadrons. The segregation of prompt and nonprompt charm hadrons can help us understand the hadronization mechanisms of both charm and beauty sectors. This can also help filter out the feed-down effects for the prompt charm measurements. In experiments, charm hadrons are reconstructed from their decay daughters, involving light-flavor hadrons ($\pi^{\pm}$, $K^{\pm}$ or $p (\bar{p})$) or leptons ($e^{\pm}$, $\mu^{\pm}$) via two or three body final state decay. As the production cross sections of charm hadrons are much smaller as compared to light-flavor hadrons, the charm hadron signal candidates are largely contaminated with combinatorial background. Therefore, several candidate selection criteria based on rectangular (fixed) cuts are imposed to reject the background candidates and improve the signal over background ratio ($S/B$). Additionally, from the signal candidates, the segregation of prompt and nonprompt charm hadrons is also a challenging task, as the production fraction of nonprompt charm hadrons is only about $\simeq 10\%$ of the inclusive charm production. The separation of prompt and nonprompt charm hadrons is mainly based on the topological features of their reconstructed decay vertices~\cite{ALICE:2024oob}. This is because the decay vertices of nonprompt charm hadrons are more displaced from the primary vertex as compared to the prompt charm hadrons. Therefore, a combination of topological variables can be used in machine learning based multi-class classification framework to precisely identify the background and signal candidates, and tag the prompt and nonprompt production modes of the charm hadrons. In recent days, efforts have been made to reconstruct and tag the prompt and nonprompt $J/\psi$ and $D^0$ mesons using machine learning models via their two body final state decay~\cite{Prasad:2023zdd, Goswami:2024xrx}. In this work, machine learning models such as XGboost and Deep Neural Network are employed for the reconstruction of $\Lambda_c^{+} (udc)$ hadrons via its three body final state decay channel, $\Lambda_c^{+} \rightarrow pK^0_s$, and $K^0_s \rightarrow \pi^{+}\pi^{-}$. This novel method performs an unbinned track-level reconstruction of the $\Lambda_c^{+}$-hadrons directly from their decay daughters, and eliminates the need of signal extraction via an invariant mass fitting method. The necessary data for this study are simulated in $pp$ collisions at $\sqrt{s}=13.6$~TeV using PYTHIA8 (Monash) model. Additionally, to mimic experiment-like conditions, momentum smearing of the final state charged hadrons and a three-dimensional smearing of the primary vertex are introduced.

The structure of the paper is as follows: Section~\ref{sec:intro} provides a brief introduction. Section~\ref{sec:method} describes the event generation using PYTHIA8 and a note on the machine learning models used in this work. The training, evaluation, and quality assurance of the models are discussed in Section~\ref{sec:MLtraining}, followed by the presentation of results and discussions in Section~\ref{sec:results}. Finally, the paper concludes with a summary of the findings in Section~\ref{sec:summary}. For the sake of simplicity, $\Lambda_c$-hadron denotes both its particle and antiparticle for the text that follows.

\section{Event Generation and Machine Learning Models}
\label{sec:method}
Event simulation with heavy-flavor production in \textit{pp} collisions at $\sqrt{s} = 13.6$~TeV is performed using the PYTHIA8 model. This model provides the necessary training dataset with the input features of the background, and the signal candidates of $\Lambda_c$-hadron. For the multi-class classification, XGBoost and Deep Neural Network models are implemented. This section briefly details the event generation using the PYTHIA8, followed by a note on the two machine learning models used in this study.

\subsection{PYTHIA8}
\label{sec:pythia}
PYTHIA8 is a pQCD-based Monte Carlo event generator which can simulate ultra-relativistic leptonic and hadronic collisions over the full range of collision energies available across experiments such as RHIC and LHC~\cite{Bierlich:2022pfr}. PYTHIA8 contains detailed physics models for simulating the evolution from a few-body hard scattering to a sophisticated multi-particle final state with special emphasis on the theory of strong interaction governed by QCD. It contains both theoretical and phenomenological models with parameters determined from data. PYTHIA8 contains several components ordered in energy (or time)-scale for the full evolution of hadronic collisions. This includes hard scattering of partons, initial and final state radiations, multiple parton interactions (MPI), treatment of beam remnants, hadronization via the Lund string fragmentation model, and particle decays and rescattering~\cite{Andersson:1983ia, Corke:2010yf}. For the heavy flavor pair production in leading-order (LO), PYTHIA8 includes two possible $2\rightarrow2$ scatterings with the mass effects. This addition of finite mass makes the matrix element expressions finite at $p_{\rm T}\rightarrow0$ limit, hence, no phase-space cuts are needed to avoid divergence~\cite{Bierlich:2022pfr}. To correctly reproduce the $D$- and $B$-meson cross sections, one needs to combine both the light and heavy flavor components of the hard-QCD processes. In this study, the default Monash 2013 tune (\texttt{Tune:pp = 14}) of PYTHIA8 (version 8.313) is used to simulate $pp$ collisions at $\sqrt{s}=13.6$~TeV~\cite{Skands:2014pea}. The charm and beauty production is included via \texttt{HardQCD:gg2ccbar = on}, \texttt{HardQCD:qqbar2ccbar = on}, \texttt{HardQCD:gg2bbbar = on}, and \texttt{HardQCD:qqbar2bbbar = on}~\cite{Combridge:1978kx}. Additionally, production of bound-states of charm and beauty quarks ($c\bar{c}$ and $b\bar{b}$) such as $J/\psi, \psi(2S), \Upsilon(nS)$ are also enabled (via \texttt{Charmonium:all=on} and \texttt{Bottomonium:all=on}). The produced $\Lambda_c^{+}$ is forced to decay into a pair of $pK_{s}^0$ via \texttt{4122:onMode = off, 4122:onIfMatch = 2212 311}. The $K_{s}^{0}$ decays to a pair of oppositely charged pions, hence, the $\Lambda_c^{+}$ are reconstructed using the three-body final state decay channel as $\Lambda_c^{+} \rightarrow pK^0_s$ (B.R. $1.59 \pm 0.07$\%) and $K^0_s \rightarrow \pi^{+}\pi^{-}$ (B.R. $69.20\pm 0.05\%$)~\cite{ALICE:2024oob}. 

\subsection{XGBoost}
Extreme Gradient Boosting or XGBoost~\cite{Chen:2016btl} is a machine learning framework based on gradient boosted decision trees. This is built upon the classification and regression tree (CART) model. Decision trees (or trees) are mathematical structures built in a top-down approach which take decisions based on certain rules giving rise to binary splitting of decision points. These points are called as nodes and the values or scores of the decision are assigned to the leaves. The tree begins from a root node and continues splitting recursively till a preset maximum depth is reached. The splitting of nodes is performed in a way that it minimizes the node impurity. In practice, many such trees are grown together, called an ensemble of trees, and the final score is the sum of scores of all the trees. For supervised models, training involves optimizing an objective function, which contains training loss and a term for regularization. In gradient boosting, the model is built in an additive forward stage-wise fashion and the addition of each new tree compensates for the shortcomings (or gradients) of the previous tree. For classification problems, the model takes a vector of features and predicts the class as the outcome. XGBoost is effective in handling large data, includes advanced features such as multi-threading and tree pruning. These features help improve the training speed. It also contains a broad range of hyperparameters that can be tuned by the user to enhance the model performance. In this work, The publicly available XGBoost v2.1.1 library~\cite{XGBoost:Online} is used with Python v3.12.9.

\subsection{Deep Neural Network}

An artificial neural network (or neural network) contains one input layer, one intermediate layer and one output layer. The input layer takes the input features, passes them through the intermediate layer to produce an output. In classification problems, the output is a class tag or label. Each layer contains active computational elements known as neurons or nodes. Input nodes simply pass the input to the next adjacent layer without performing any mathematical operation. For passing of information between layers, several or all nodes of two consecutive layers are interconnected. If there are more than one intermediate layer, then it is called a deep neural network. Similarly, if all the nodes of two adjacent layers are connected, it forms a dense layer. Each such connection has some weight and each layer has its own bias. The smallest computing element of a neural network is the perceptron. Each perceptron performs two tasks. First, it collects the values from the connected nodes of it previous layer multiplied with the weights, and second, it calculates the output value based on a nonlinear transformation or activation. After the calculation of several such perceptrons, the final output is compared to the actual value of the output. The accuracy of the model is evaluated through a suitable loss function. The objective is to optimize the weights and biases of the model such that the loss function is minimized. In the forward cycle, input values flow through the network, and the loss function is computed. In the back-propagation cycle, the model optimizes the weights and biases. This process continues through the entire dataset iteratively, taking a small batch of input values. Each complete cycle is called an epoch. The training stops when some convergence criteria is achieved. For the implementation of the DNN model, publicly available KERAS v3.6.0~\cite{Keras:Docs} deep learning
Application Programming Interface (API) with TensorFlow v2.17.0~\cite{tensorflow:Docs} has
been used in this work. The code is developed in Python v3.12.9 with the help of the
scikit-learn framework~\cite{Scikit-Learn:Docs}. For simplicity, hereafter, XGBoost and Deep Neural Network are termed as XGB and DNN, respectively.

\section{Training and Evaluation}
\label{sec:MLtraining}
This section describes the chosen input features, the parameter settings of the machine learning models, training, evaluation, and quality assurance.

\subsection{Input features}

\begin{figure*}[ht!]
    \centering
        \includegraphics[width=0.32\textwidth]{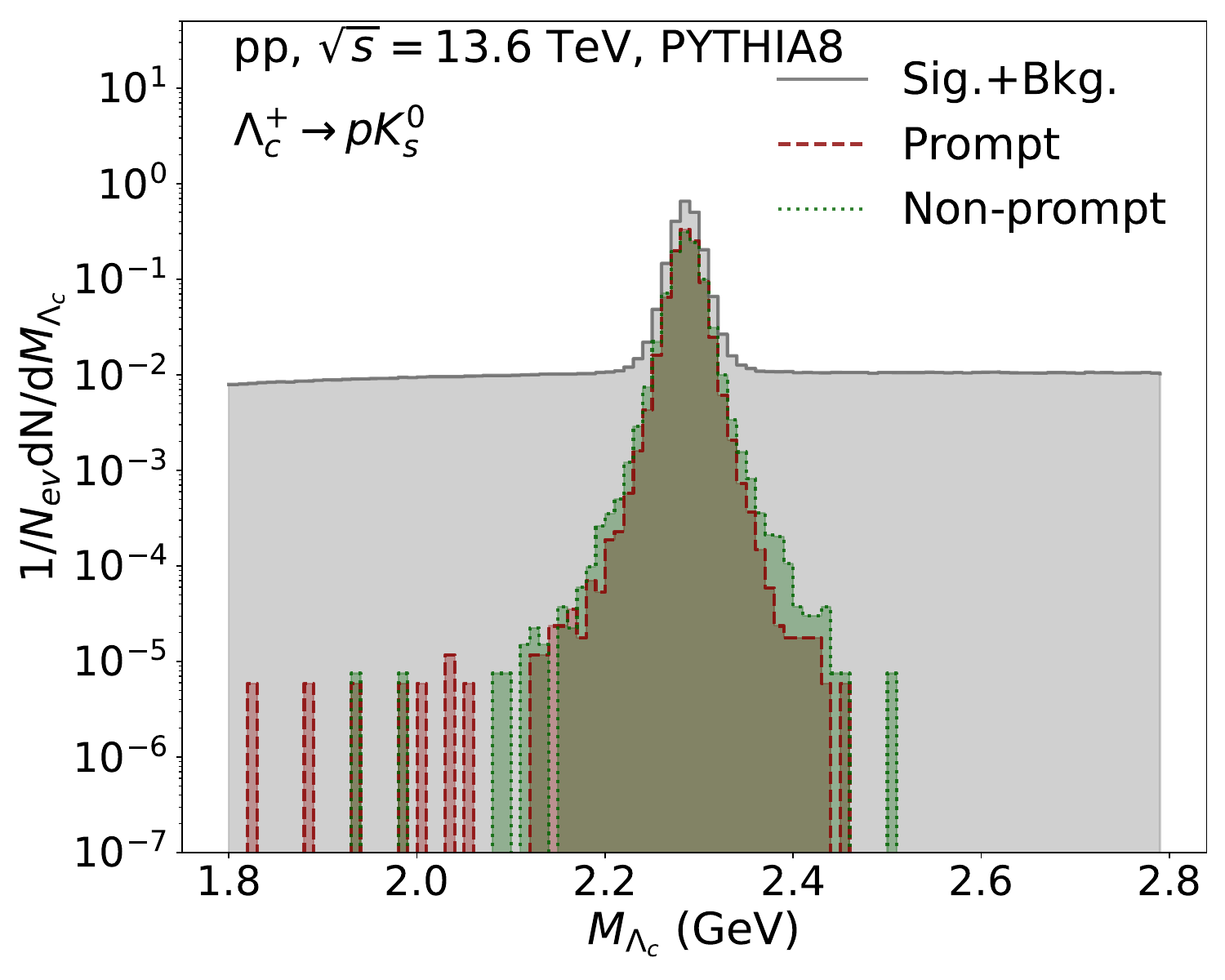}
        \includegraphics[width=0.32\textwidth]{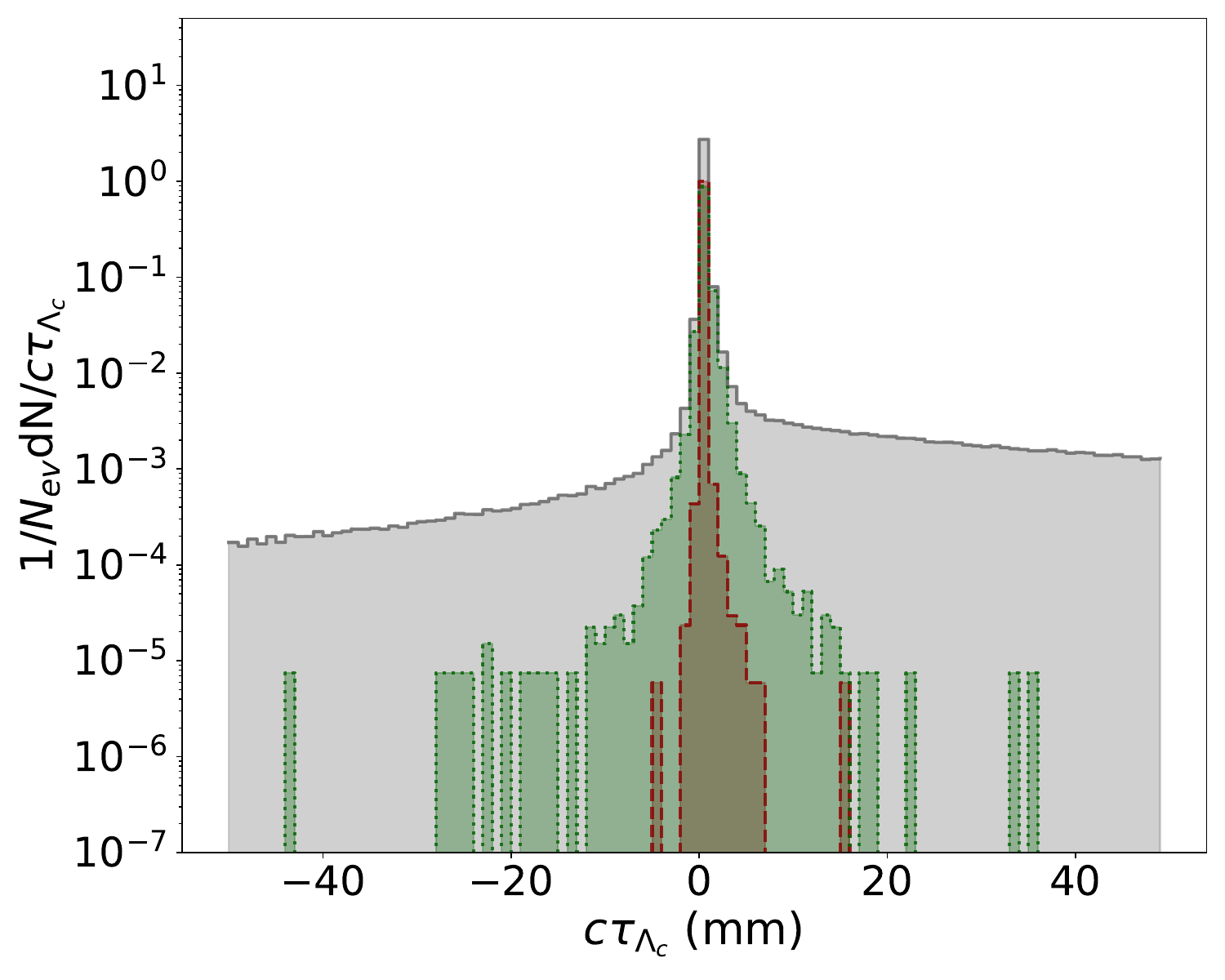}
        \includegraphics[width=0.32\textwidth]{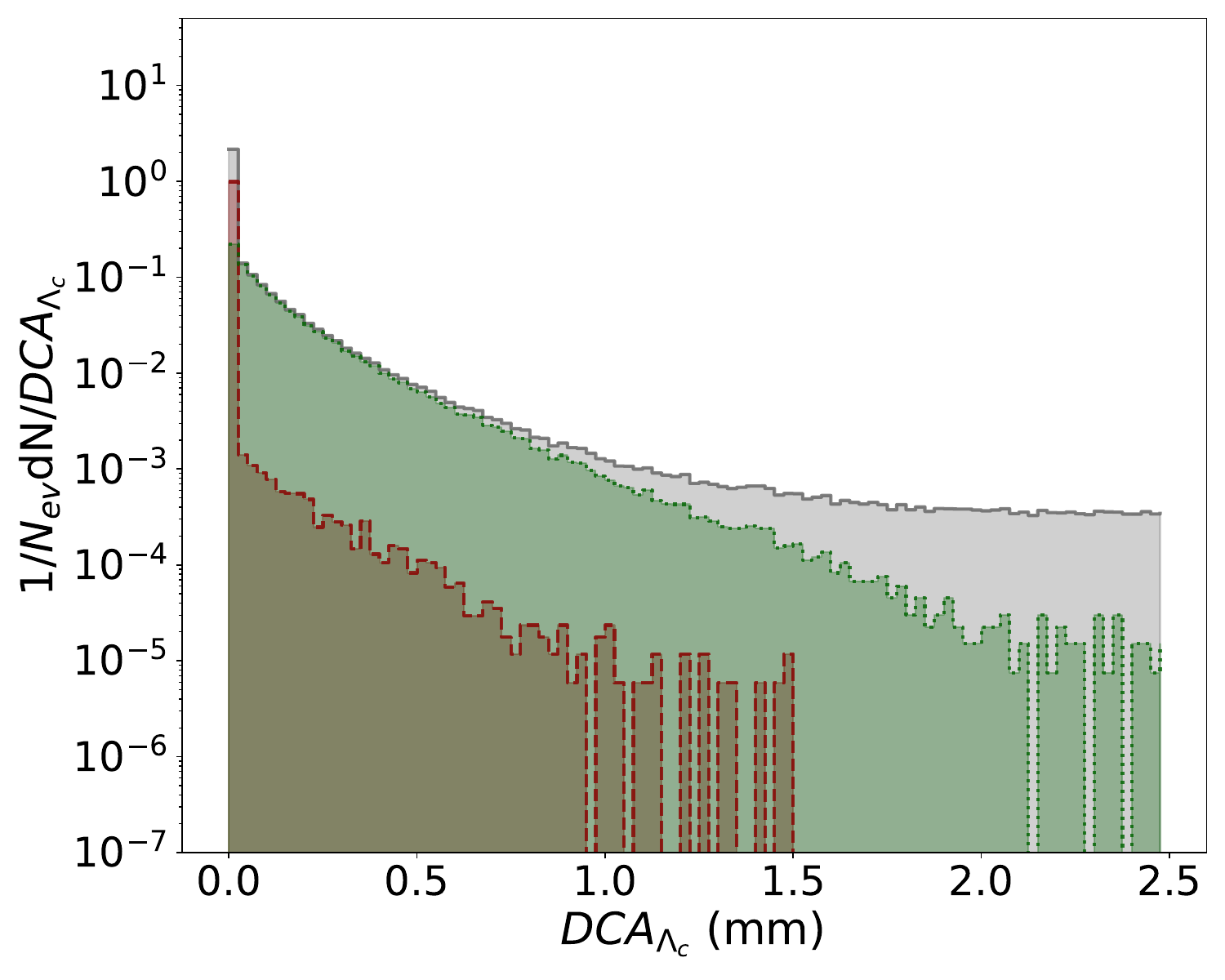}
        \includegraphics[width=0.32\textwidth]{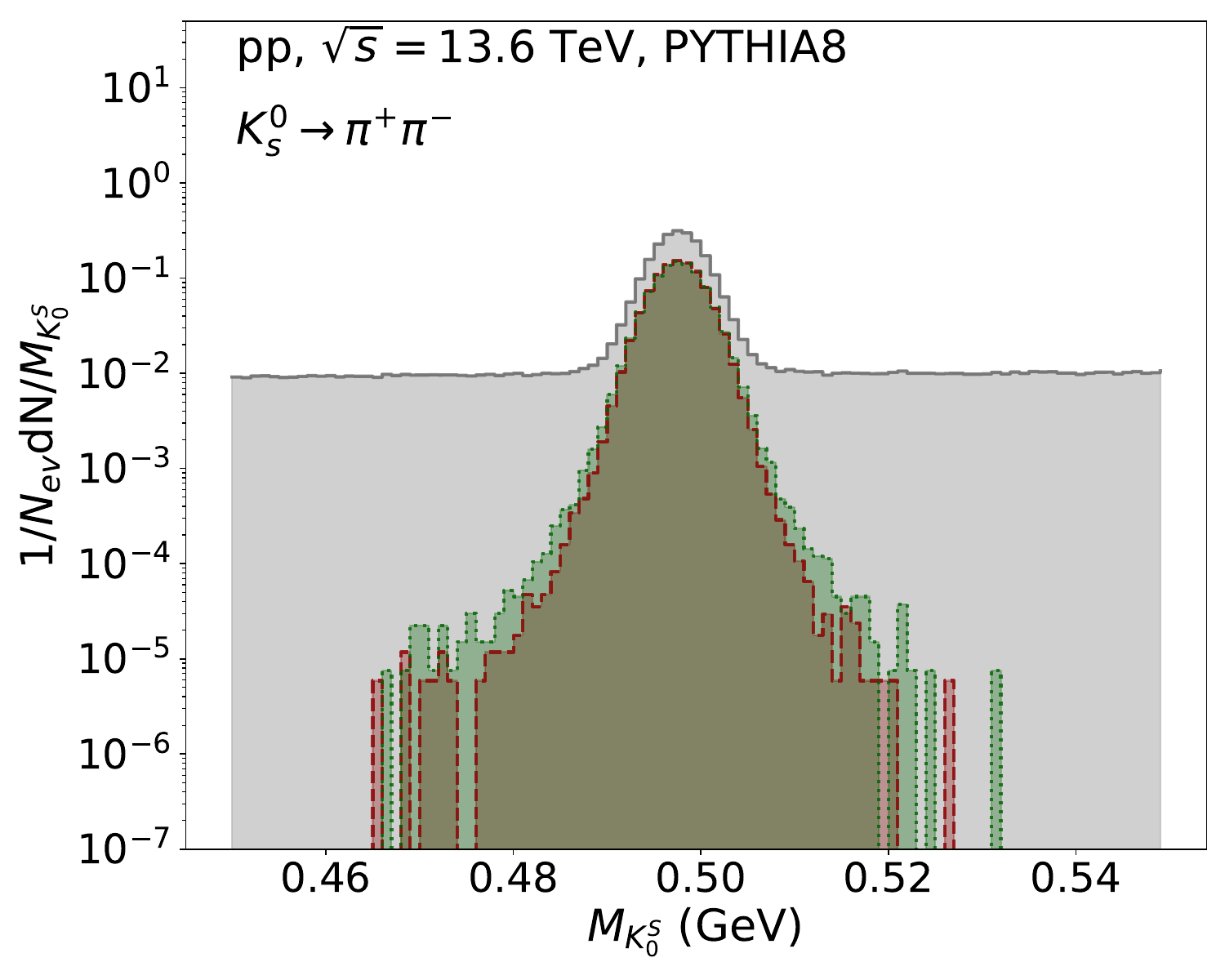}
        \includegraphics[width=0.32\textwidth]{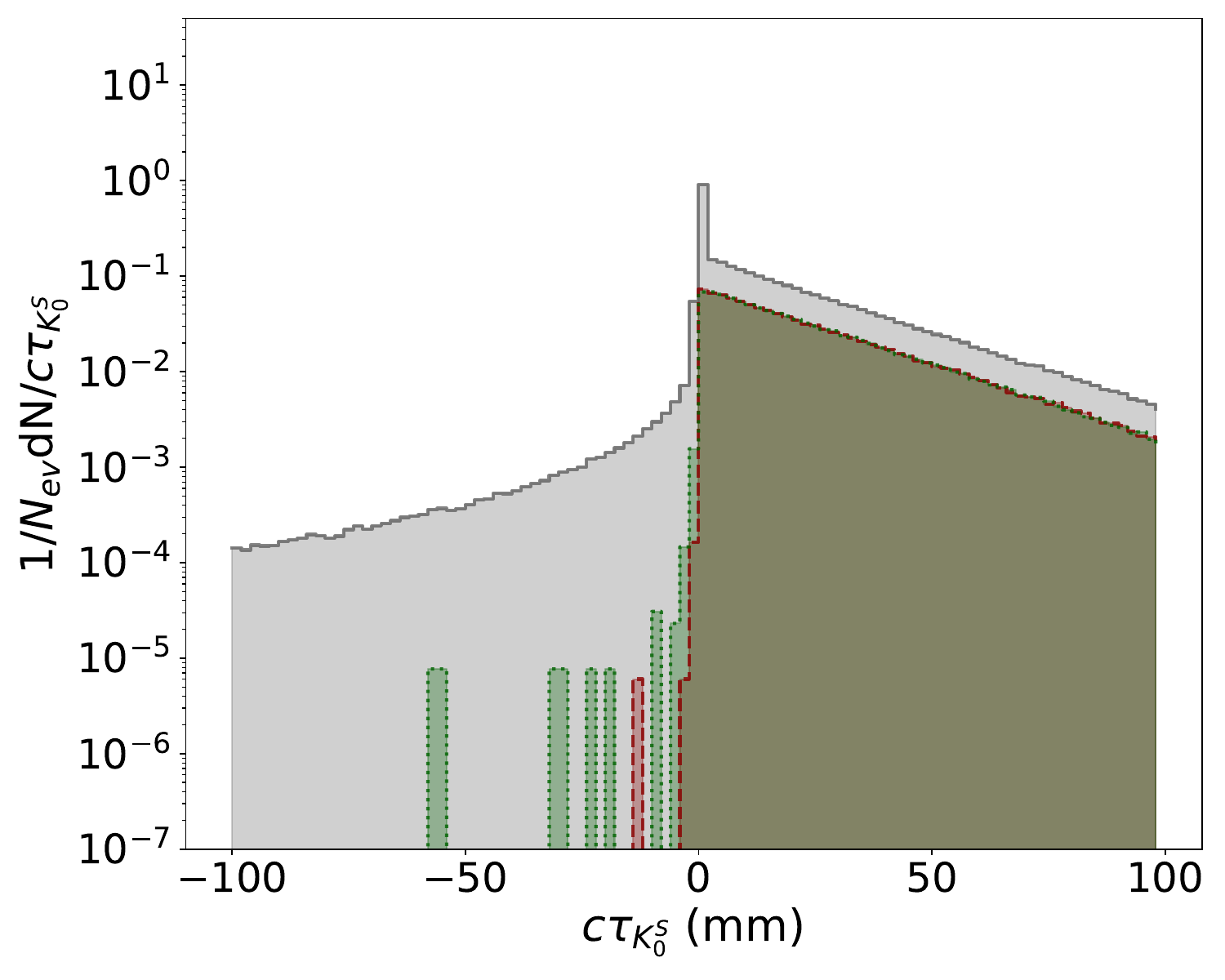}
        \includegraphics[width=0.32\textwidth]{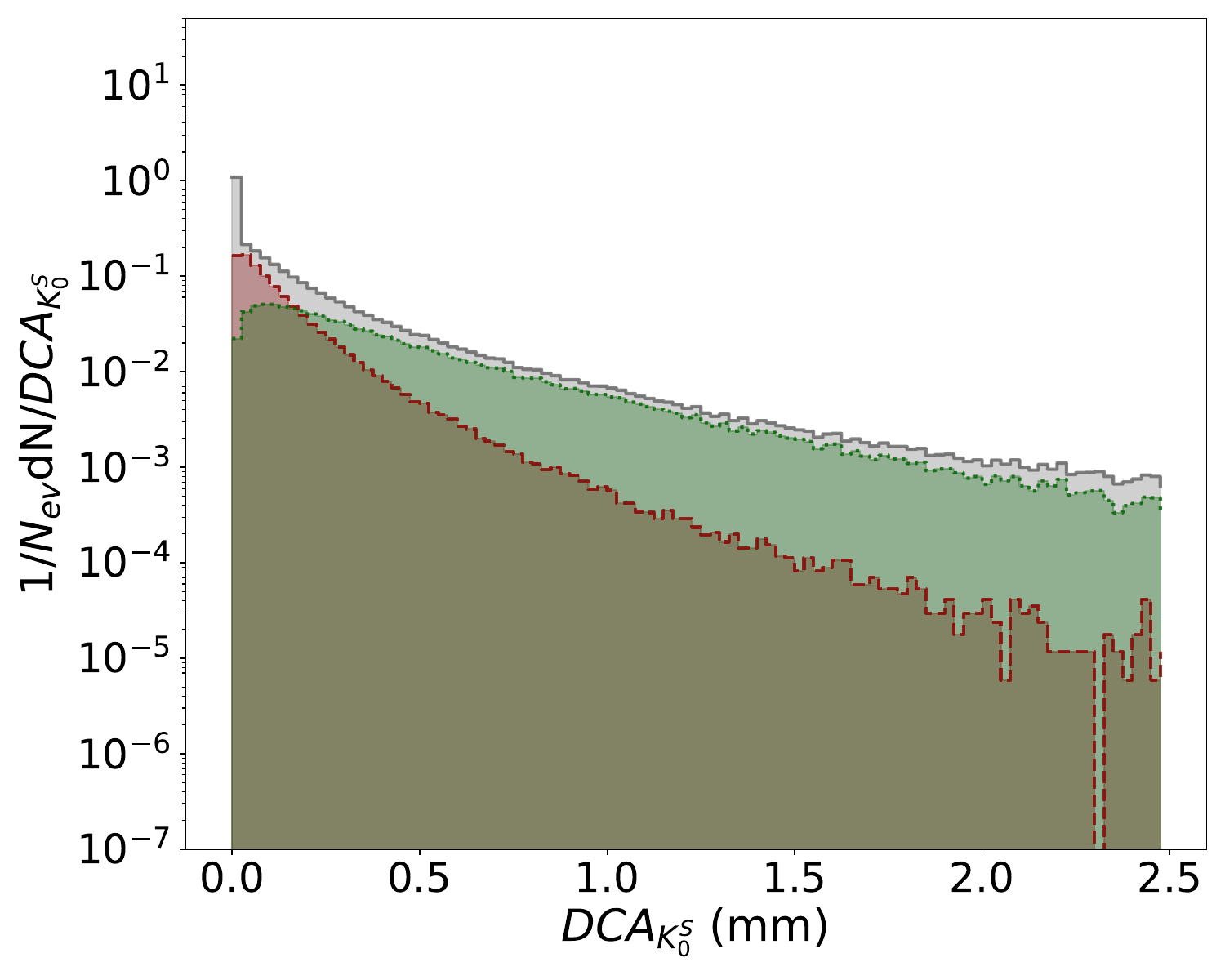}
    \caption{Distribution of the features used in this analysis. The distributions are event normalized. The gray histograms show the sum of signal and background,  while the red dashed and green dotted lines represent the prompt and non-prompt \ensuremath{\Lambda_c^+} signal components, respectively.}
    \label{fig:features}
\end{figure*}

From the three body final state decay of $\Lambda_c$ hadron, its invariant mass ($M_{\rm inv.}^{\Lambda_c}$), pseudoproper decay length ($c\tau_{\Lambda_{c}}$), and the distance of closest approach (${\rm DCA}_{\Lambda_c}$) are used as the first set of input features. In addition, the invariant mass of $K^{0}_{s}$ decaying through a pair of oppositely charged pions ($M_{\rm inv.}^{K^{0}_{s}}$), its pseudoproper decay length ($c\tau_{K^{0}_{s}}$) and the distance of closest approach (${\rm DCA}_{K^{0}_s}$) are taken as secondary input features. The primary input features are well suited for signal and background identification, whereas the addition of secondary features, helps in resolving the topological separation, \textit{i.e.,} the prompt and nonprompt $\Lambda_c^{+}$. For each combination of the three decay daughters, if both the pions have the same mother as $K_s^0$ and the $K_s^0$ and $p$ (or $\bar{p}$) have the same mother as $\Lambda_c^{+}$, then it is tagged as a signal; else, it is tagged as background. For each signal candidate, further checks are performed based on the list of mothers of the $\Lambda_c^{+}$ mentioned in Tab.~\ref{tab:heavy_baryons} to tag its topological production mode, which can be a prompt or nonprompt candidate. The tags, $0$, $1$, and $2$ are assigned for background, prompt, and nonprompt combinations, respectively. To reduce memory usage, only the combinations with invariant mass, $1.8<M_{\rm inv.}^{\Lambda_c}<2.8$ GeV and $0.45<M_{\rm inv.}^{K^{0}_{s}}<0.55$ GeV, are selected. The standard definitions of the pseudoproper decay length and distance of the closest approach are given below.

\begin{equation}
    c\tau = \frac{c~M_{\rm inv.}~\vec{L} . \vec{p_{\rm T}}}{|\vec{p_{\rm T}}|^2}
\end{equation}

\begin{equation}
    {\rm DCA} = |\vec{L}|.\sin \theta
\end{equation}

Here, $\vec{L}$ is the vector from the primary vertex to the decay vertex of the particle, such as $\Lambda_c$ or $K^{0}_{s}$, and $c=1$ is the velocity of light in natural units. The pointing angle $\theta$ is the angle made by the momentum vector of the particle, $\vec{p}$ with $\vec{L}$, given as, $\theta = \cos^{-1}\Big(\frac{\vec{p}.\vec{L}}{|\vec{p}||\vec{L}|}\Big)$. All these input features are also available in experiments.

\begin{table}[ht!]
    \centering
    \caption{List of charm and bottom hadrons that can produce $\Lambda_c$ with their PDG identification numbers and quark content.}
    \begin{tabular}{lll}
        \hline
        \textbf{PDG ID} & \textbf{Particle} & \textbf{Quark Content} \\
        \hline
        \multicolumn{3}{c}{\textbf{Charm Hadrons}} \\
        \hline
        4112 & $\Sigma_c^0$ & $ddc$ \\
        4114 & $\Sigma_c^{*0}$ & $ddc$ \\
        4124 & $\Lambda_c(2625)^+$ & $udc$ \\
        4212 & $\Sigma_c^+$ & $udc$ \\
        4214 & $\Sigma_c^{*+}$ & $udc$ \\
        4222 & $\Sigma_c^{++}$ & $uuc$ \\
        4224 & $\Sigma_c^{*++}$ & $uuc$ \\
        14122 & $\Lambda_c(2595)^+$ & $udc$ \\
        \hline
        \multicolumn{3}{c}{\textbf{Bottom Hadrons}} \\
        \hline
        511 & $B^0$ & $\bar{d}b$ \\
        521 & $B^+$ & $\bar{u}b$ \\
        531 & $B_s^0$ & $\bar{s}b$ \\
        541 & $B_c^+$ & $\bar{c}b$ \\
        545 & $B_c^{*+}$ & $\bar{c}b$ \\
        5122 & $\Lambda_b^0$ & $udb$ \\
        5132 & $\Xi_b^-$ & $dsb$ \\
        5232 & $\Xi_b^0$ & $usb$ \\
        5332 & $\Omega_b^-$ & $ssb$ \\
        \hline
    \end{tabular}
    \label{tab:heavy_baryons}
\end{table}

\begin{table}[ht!]
\centering
\caption{Parameters and their values for the beam vertex spread in PYTHIA8~\cite{ALICE:2010vtz}.}
\begin{tabular}{cc}
\hline
PYTHIA8 String     & Value (in mm) \\ \hline
\texttt{Beams:allowVertexSpread} & on    \\
\texttt{Beams:sigmaVertexX}      & 0.23   \\
\texttt{Beams:sigmaVertexY}      & 0.27  \\
\texttt{Beams:sigmaVertexZ}      & 40.24 \\
\texttt{Beams:maxDevVertex}      & 5     \\
\texttt{Beams:offsetVertexX}     & -0.35 \\
\texttt{Beams:offsetVertexY}     & 1.63  \\
\texttt{Beams:offsetVertexZ}     & -4.0  \\ \hline
\end{tabular}
\label{tab:vtxSpread}
\end{table}

To replicate experimental conditions, the spread of the beam interaction vertex and the smearing of the transverse momentum ($p_{\rm T}$) are introduced. In experiments, the beam interaction vertex has a three-dimensional spread, which means, the primary vertex follows a Gaussian distribution around the global origin $ = (0,0,0)$. In PYTHIA8, this spread of the interaction vertex can be simulated according to a simple Gaussian distribution. The offset and sigma of the spread of the vertices in each of the Cartesian axes are taken from~\cite{ALICE:2010vtz} and are also mentioned in Tab.~\ref{tab:vtxSpread}. The implementation of the $p_{\rm T}$ smearing depends on the $p_{\rm T}$ resolution of the detector, which can be written as,
\begin{equation}
    \big(\sigma(p_{\rm T})/p_{\rm T}\big)^{2} = a^2 + \big(b p_{\rm T}\big)^2. 
    \label{eq:pTSmearing}
\end{equation}
Here, the momentum dependent $p_{\rm T}$ resolution is $\sigma(p_{\rm T})$. The values of the constant term $a=0.01$ and the stochastic term $b=0.007$ are obtained from the ALICE experiment~\cite{ALICE:2010vtz}. Using the $p_{\rm T}$ resolution given in Eq.~\eqref{eq:pTSmearing}, $p_{\rm T}$ smearing is obtained using a Gaussian distribution with width $\sigma(p_{\rm T})$, and applied to all the final-state tracks. Additionally, a cut in the $z$-vertex, as $|V_{z}|<10$ cm, is applied to select good events according to the event selection criteria of the ALICE experiment.

Figure~\ref{fig:features} shows the distribution of the features used in this analysis. The distributions are normalized by the total number of events. The gray histograms show the sum of the signal and background,  while the red dashed and green dotted lines represent the prompt and nonprompt \ensuremath{\Lambda_c} signal components, respectively. Due to the $p_{\rm T}$ dependent smearing of the tracks, the invariant mass distributions could generate a finite width around the true invariant mass of the mother particle which is observed in experiments, rather than a sharp delta-like peak. Since, the nonprompt $\Lambda_c$ are produced from the beauty-hadrons, they can decay at a larger distance than the prompt ones, resulting in a broader $c\tau_{\Lambda_c}$ and ${\rm DCA}_{\Lambda_c}$ distribution as seen in the figure. A similar effect for nonprompt candidates can also be seen in the ${\rm DCA}_{K^0_s}$ distribution.


\subsection{Model training}

\begin{figure}[ht!]
    \centering
        \includegraphics[width=0.45\textwidth]{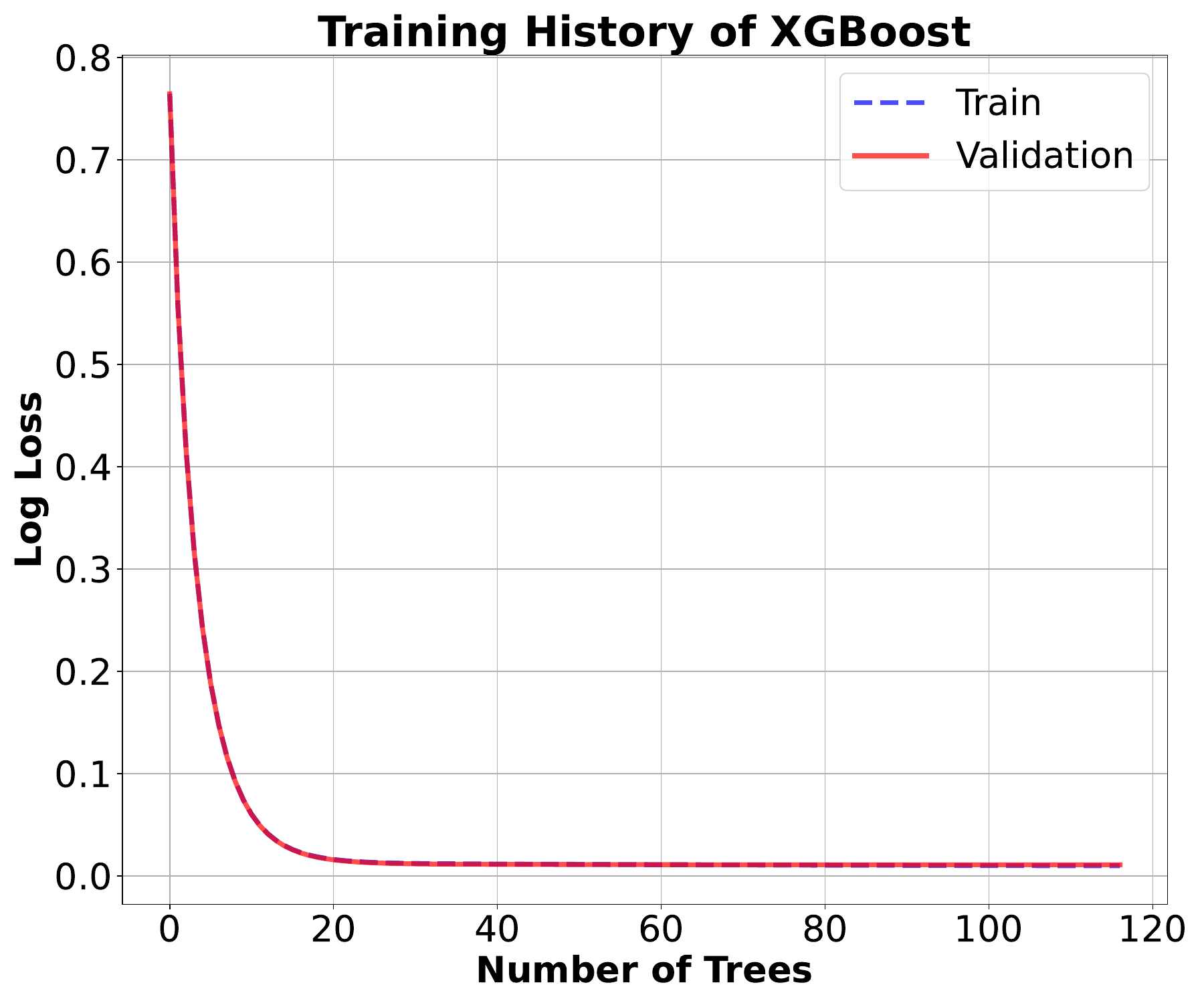}
        \includegraphics[width=0.45\textwidth]{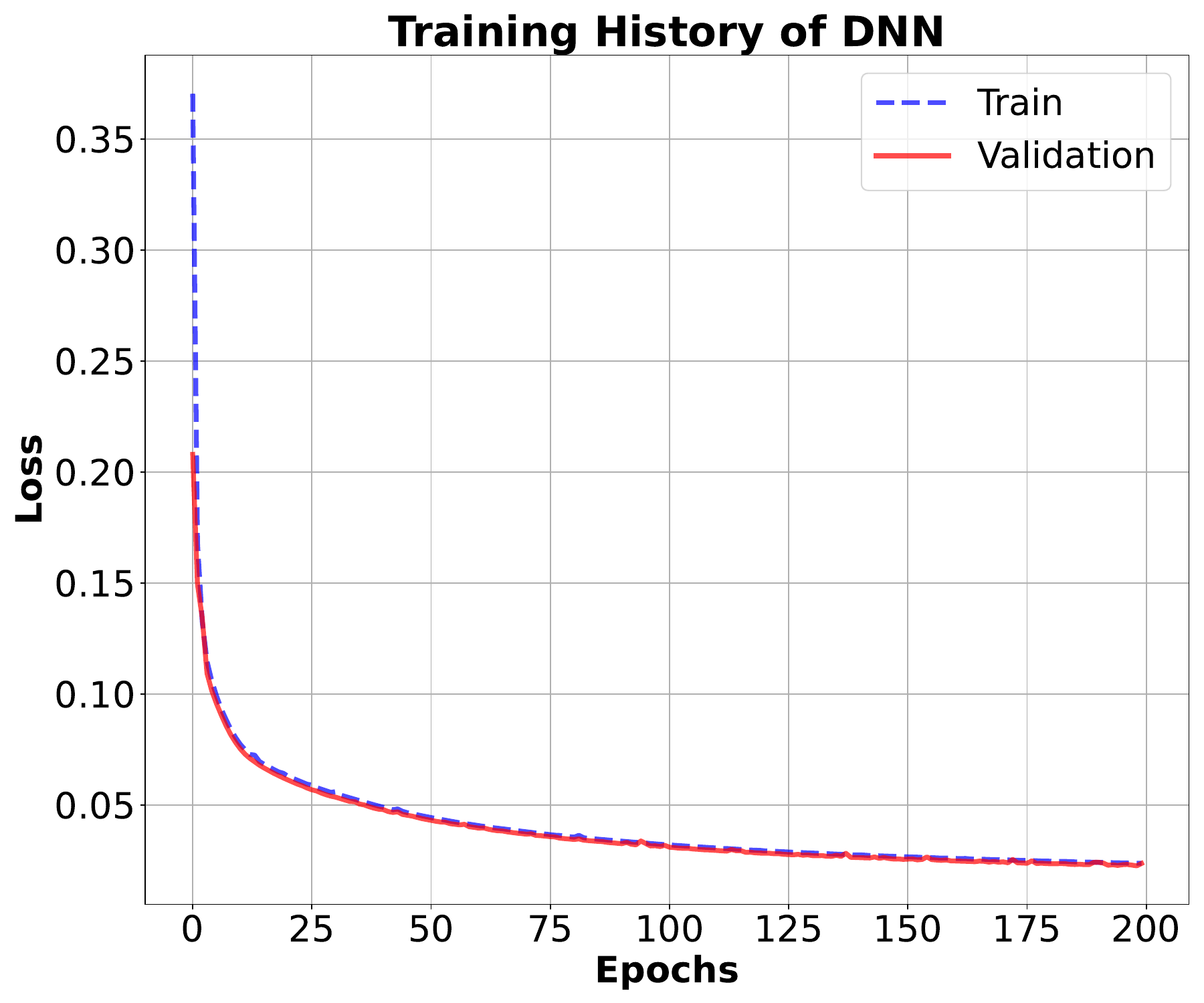}
    \caption{Evolution of the loss functions on the training and validation sample during the model training as a function of number of trees for the XGB (top) and epochs for the DNN (bottom).}
    \label{fig:TrainingCurve}
\end{figure}

For the training of models, each combination of $\Lambda_c$ candidates with all the input features shown in Fig.~\ref{fig:features} and its corresponding tag, (\textit{i.e.,} 0, 1 or 2) are fed into the model. As this is a multiclass classification problem, ideally, one should train the model with an equal pool of samples from each class so that the model is not biased. However, in general, only a small fraction of all such combinations could actually be the signal; therefore, the background sample size can grow larger compared to the signal, making the background a majority class. To address this issue, only 1\% of the total combinatorial background is randomly selected to construct the background sample pool. The sample of signal candidates can also contain prompt and non-prompt candidates. Again, the production fraction of the nonprompt candidates is much lower than that of the prompt ones. This is because the production of nonprompt candidate necessitates the production of beauty quarks from the initial hard scatterings and the production cross-section for beauty quarks is less than that of the charm quarks. Therefore, there could be a natural class imbalance between the prompt and nonprompt candidates in the training sample pool obtained from a minimum bias dataset. One practical way to increase the overall signal including the prompt and nonprompt candidates in simulation is to selectively enable the processes that can produce abundant charm and beauty quarks. This is possible in simulations such as in PYTHIA8 via enabling only the $gg\rightarrow c\bar{c}$, $q\bar{q} \rightarrow c\bar{c}$, $gg\rightarrow b\bar{b}$, and $q\bar{q} \rightarrow b\bar{b}$ hard QCD processes. Such a simulation is biased in a way that it contains at least one pair of charm or beauty quarks produced in every hadronic interaction, which may not be the case in a minimum bias collision, and therefore, termed as the charm and beauty enhanced production. However, this serves the role of equalizing the prompt and nonprompt sample size and overall improvement in the number of signal candidates. This enhancement of charm and beauty production does not affect its kinematic or topological production mechanism, therefore, it can be used as a training instance. It is to be noted here that, several sampling techniques based on the creation of synthetic data are available to equalize the sample size for each class. This is done by either oversampling the minority class or undersampling the majority class. One has to use these methods very carefully as they can dilute the input-output correlation while creating the synthetic data, and it may result in reducing the overall reliability of the model for another independent set of samples.

\begin{table}[ht!]
\centering
\caption{XGBoost Model Hyperparameters}
\begin{tabular}{ll}
\hline
\textbf{Parameter} & \textbf{Value} \\
\hline
Booster & \texttt{gbtree} \\
Learning Rate & 0.25 \\
Number of Estimators (max.) & 1000 \\
Maximum Tree Depth & 6 \\
Objective Function & \texttt{multi:softmax} \\
Evaluation Metric & \texttt{mlogloss} \\
Number of Classes & 3 \\
Early Stopping Rounds & 10 \\
\hline
\end{tabular}
\label{tab:xgb_params}
\end{table}

\begin{table}[htbp]
\centering
\caption{DNN Architecture and Training Parameters}
\begin{tabular}{ll}
\hline
\textbf{Parameter} & \textbf{Value} \\
\hline
\multicolumn{2}{l}{\textit{I. Architecture}} \\
Input Layer & 8 neurons \\
Hidden Layer 1 & 128 neurons (\texttt{ReLU}) \\
Hidden Layer 2 & 256 neurons (\texttt{ReLU}) \\
Hidden Layer 3 & 128 neurons (\texttt{ReLU}) \\
Output Layer & 3 neurons (\texttt{Softmax}) \\
\hline
\multicolumn{2}{l}{\textit{II. Training Parameters}} \\
Optimizer & \texttt{Adam} \\
Learning Rate (LR) & 1$\times$10$^{-5}$ \\
Loss Function & \texttt{Sparse Categorical} \\ & \texttt{Cross-entropy} \\
Batch Size & 2048 \\
Maximum Epochs & 200 \\
\hline
\textit{III. Call--backs} & \\
Early Stopping Patience & 5 \\
LR Reduction Factor & 0.1 \\
LR Patience & 10 \\
Minimum LR & 1$\times$10$^{-6}$ \\
\hline
\end{tabular}
\label{tab:dnn_params}
\end{table}

From the entire sample, 90\% of the instances are randomly selected for the training, and the remaining 10\% instances are kept for the testing. From the training sample and additional 10\% instances are allocated for the validation. The used configurations for the XGB and DNN models are listed in Tab.~\ref{tab:xgb_params} and \ref{tab:dnn_params}, respectively.. 

\subsection{Quality assurance}
\label{sec:qa}

The quality of the model training can be evaluated by studying the evolution of the loss function. The loss function gives a numerical estimation of the difference between the true and the model predicted class for a given set of input features. For an ideal model, the loss function should approach to zero, which means the true and predicted classes are exactly identical. In this study, during the training of the models, the evolution of loss functions on the training and validation sample are recorded and are shown in Fig.~\ref{fig:TrainingCurve} as a function of number of trees and epochs for the XGB and DNN models, respectively. The loss function starts from a higher value in the beginning of the training, and sharply falls to a much lower value as the training progresses further. It continues to decrease smoothly until a saturation behavior occurs, which means the value of the loss does not change much even if the training continues. After a certain number of trees and epochs, the training is stopped using the early stopping call--backs, which ensures minimal overfitting. The evolution of the loss function in the training and validation sample yields a similar outcome, which means, the model is not biased towards the training sample.

\begin{figure}[ht!]
    \centering
        \includegraphics[width=0.45\textwidth]{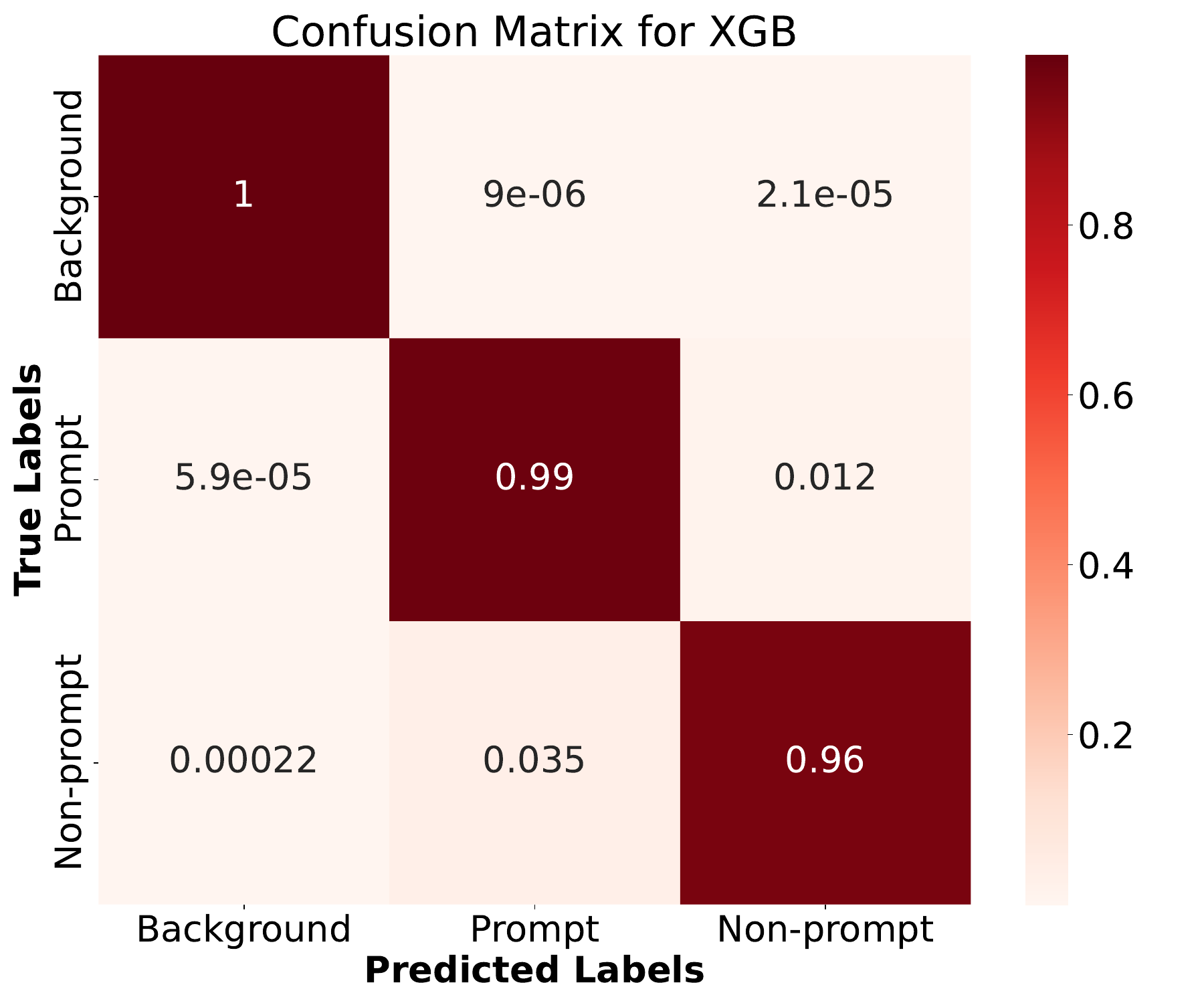}
        \includegraphics[width=0.45\textwidth]{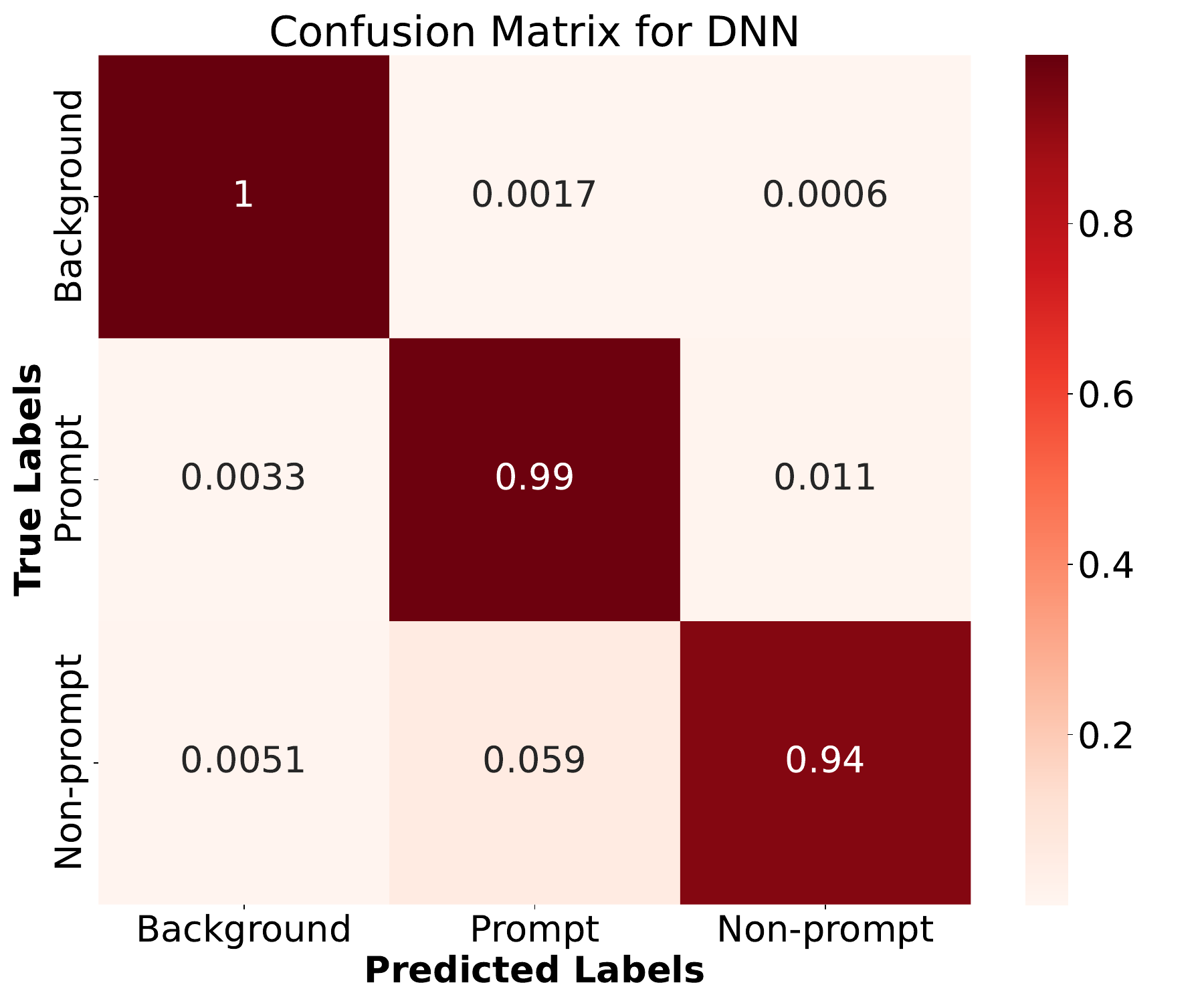}
    \caption{Confusion matrix from XGB (top) and DNN (bottom) depicting the normalized fraction of the prediction and true tags for background, prompt, and nonprompt candidates.}
    \label{fig:ConfMatrix}
\end{figure}

\begin{figure}[ht!]
    \centering
        \includegraphics[width=0.48\textwidth]{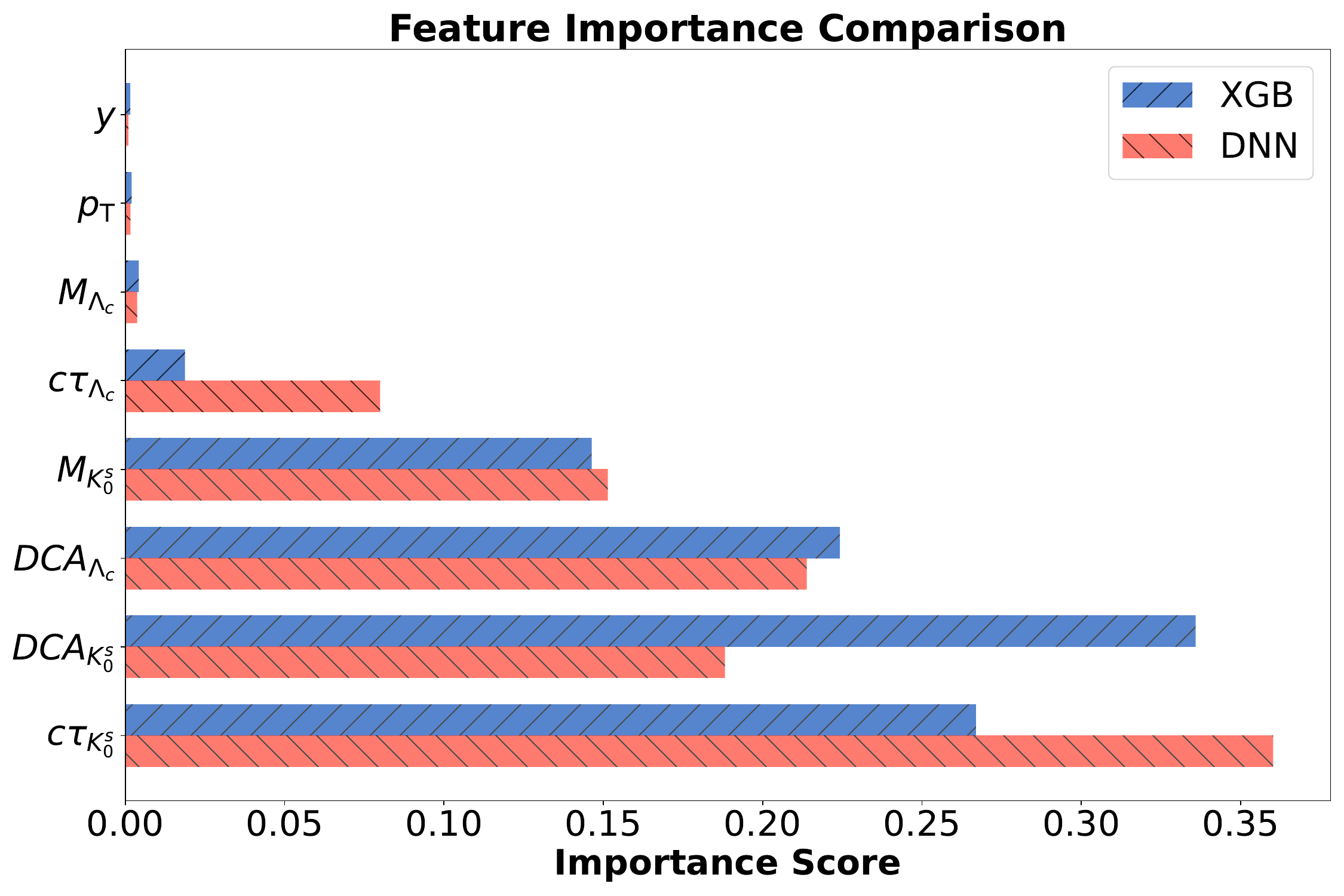}
    \caption{Normalized feature importance score obtained from the XGB (blue) and the DNN (red) depicting the importance of a certain feature in classifying the candidates.}
    \label{fig:featureImp}
\end{figure}

Figure~\ref{fig:ConfMatrix} presents the normalized confusion matrices for both XGB and DNN models. The rows and columns correspond to the true and predicted labels for the three classes in the sample, \textit{i.e.,} background, prompt, and nonprompt. Each element $(i,j)$ in these matrices represents the fraction of instances from true class $i$ that were classified as class $j$ by the models. The diagonal elements indicate correct classifications, while the off-diagonal elements represent misclassifications. A darker color corresponds to a higher fraction, which populates the diagonal elements. This indicates the model's accuracy in classifying the respective classes correctly.
A perfect classifier would show values of 1.0 along the diagonal and 0.0 elsewhere, while any deviations from this would indicate classification errors between the respective classes. Both the models can identify the background candidates over the signal with almost 100\% accuracy. Again, both the models succeed to identify the prompt candidates with almost 99\% accuracy. The XGB model has almost 96\% accuracy for the nonprompt candidate tagging whereas the DNN model can only reach up to 94\% for the same class. That leads to a 3.5\% and 6\% misidentification rate for the nonprompt candidates as prompt candidates from the XGB and DNN models, respectively. With the current model configurations, the XGB model slightly outperforms the DNN model when the nonprompt tagging is concerned, while keeping a similar accuracy for the prompt and background candidates tagging. The classification errors for the XGB model is also less as compared to the DNN model.

Figure~\ref{fig:featureImp} shows the normalized feature importance score assigned to each input feature evaluated from the XGB (blue) and the DNN (red) models. The features are sorted in ascending order of the mean importance of the two models. For the XGB model, the importance score is associated with the number of times a certain feature is used to split a node. More important features are used repeatedly in node splitting, and hence, they are assigned a higher score. Similarly, for the DNN model, the importance score for a certain feature is calculated by shuffling its value to a random number, and then checking its effect on the prediction accuracy. If the feature is important for the model, this should reduce the accuracy. For the XGB model, the feature importance ranking is ${\rm DCA}_{k^0_s} > c\tau_{k^0_s} > {\rm DCA}_{\Lambda_c} > M_{k^0_s} > c\tau_{\Lambda_c}> M_{\Lambda_c}$, while for the DNN model, it is $ c\tau_{k^0_s} > {\rm DCA}_{\Lambda_c} > {\rm DCA}_{k^0_s} > M_{k^0_s} > c\tau_{\Lambda_c}> M_{\Lambda_c}$.

\section{Results and discussions}
\label{sec:results}

As mentioned earlier, simulations with charm and beauty enhanced production yield more $\Lambda_c$ hadrons compared to the minimum bias collisions. While this enhanced dataset is suitable for training the models, practical applications require simulating minimum bias events with all possible hardQCD processes enabled. This approach produces charm and beauty quarks as per their interaction cross-sections and can be used to derive physics for the $\Lambda_c$ production. Therefore, all the results presented in this Section, are obtained from a minimum bias $pp$ collisions dataset using PYTHIA8 model (Monash).

\begin{figure}[ht!]
    \centering
        \includegraphics[width=0.45\textwidth]{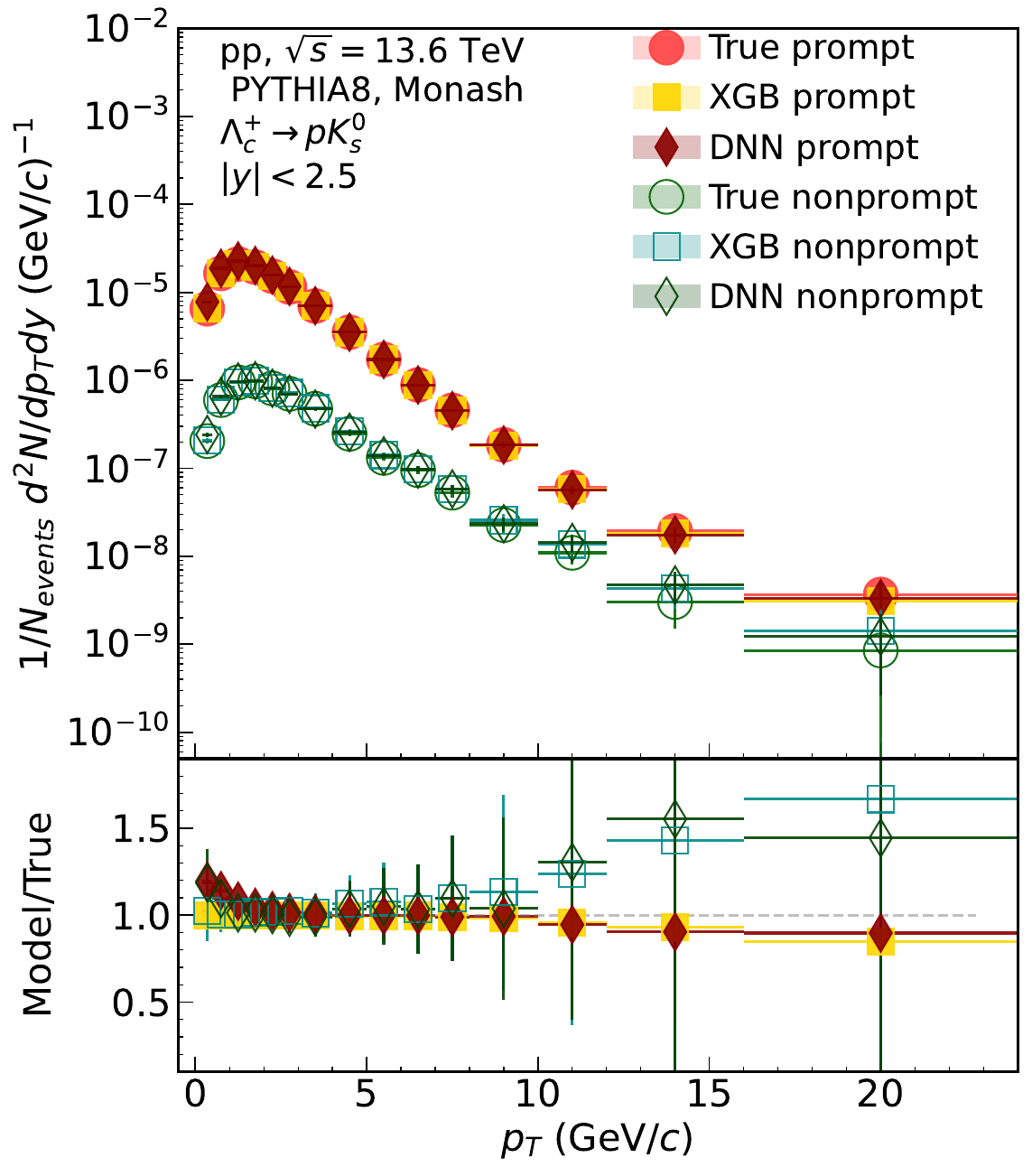}
    \caption{Transverse momentum distribution of prompt and nonprompt $\Lambda_c$-hadron in minimum bias $pp$ collisions at $\sqrt{s} = 13.6$~TeV in $|y|<2.5$, from the PYTHIA8 model along with the predictions from the XGB and DNN models. The bottom panel shows the ratio between predicted and true values for both XGB and DNN models.}
    \label{fig:pTSpectra}
\end{figure}

Figure~\ref{fig:pTSpectra} shows the transverse momentum distribution of prompt and nonprompt $\Lambda_c$ in minimum bias $pp$ collisions at $\sqrt{s} = 13.6$~TeV in $|y|<2.5$, from the PYTHIA8 model along with the predictions from the XGB and DNN models. The bottom panel shows the ratio between predicted and true values for both XGB and DNN models. The $p_{\rm T}$ spectra are reconstructed from $\Lambda_{c}^{+}\rightarrow pK^{0}_{s}$ channel. In low-$p_{\rm T}$, the yield of nonprompt $\Lambda_c$ is almost one order of magntitude lesser than that of the prompt $\Lambda_c^{+}$. However, this difference between the nonprompt and prompt yield gradually decreases towards high-$p_{\rm T}$, even approaching almost similar values beyond $p_{\rm T} \geq 16$~GeV/$c$. This is due to the fact that the production of nonprompt $\Lambda_c^{+}$ comes from the charm quark that has been produced from the beauty hadron decay, and the production of charm quarks from beauty hadrons favors higher transverse momenta. Similar observations for other charm hadrons such as $D^0$ and $J/\psi$ have also been observed~\cite{Prasad:2023zdd, Goswami:2024xrx}, which supports this argument. The pre-trained XGB and DNN models are then used to distinguish the $\Lambda_c$-hadrons candidates from the combinatorial background using the input features described in Section~\ref{sec:MLtraining}. From the signal candidates, the models identify the prompt and nonprompt $\Lambda_c$-hadrons. Both the ML models can successfully tag the signal candidates and their topological production modes. However, the XGB model outperforms the DNN model. This can also be seen from the bottom ratio plots, where the ratio of XGB to the true remains almost flat at unity showcasing a better agreement with the true. The ratio between the DNN and true curves have fluctuations around unity, especially at low-$p_{\rm T}$.

\begin{figure}[ht!]
    \centering
        \includegraphics[width=0.45\textwidth]{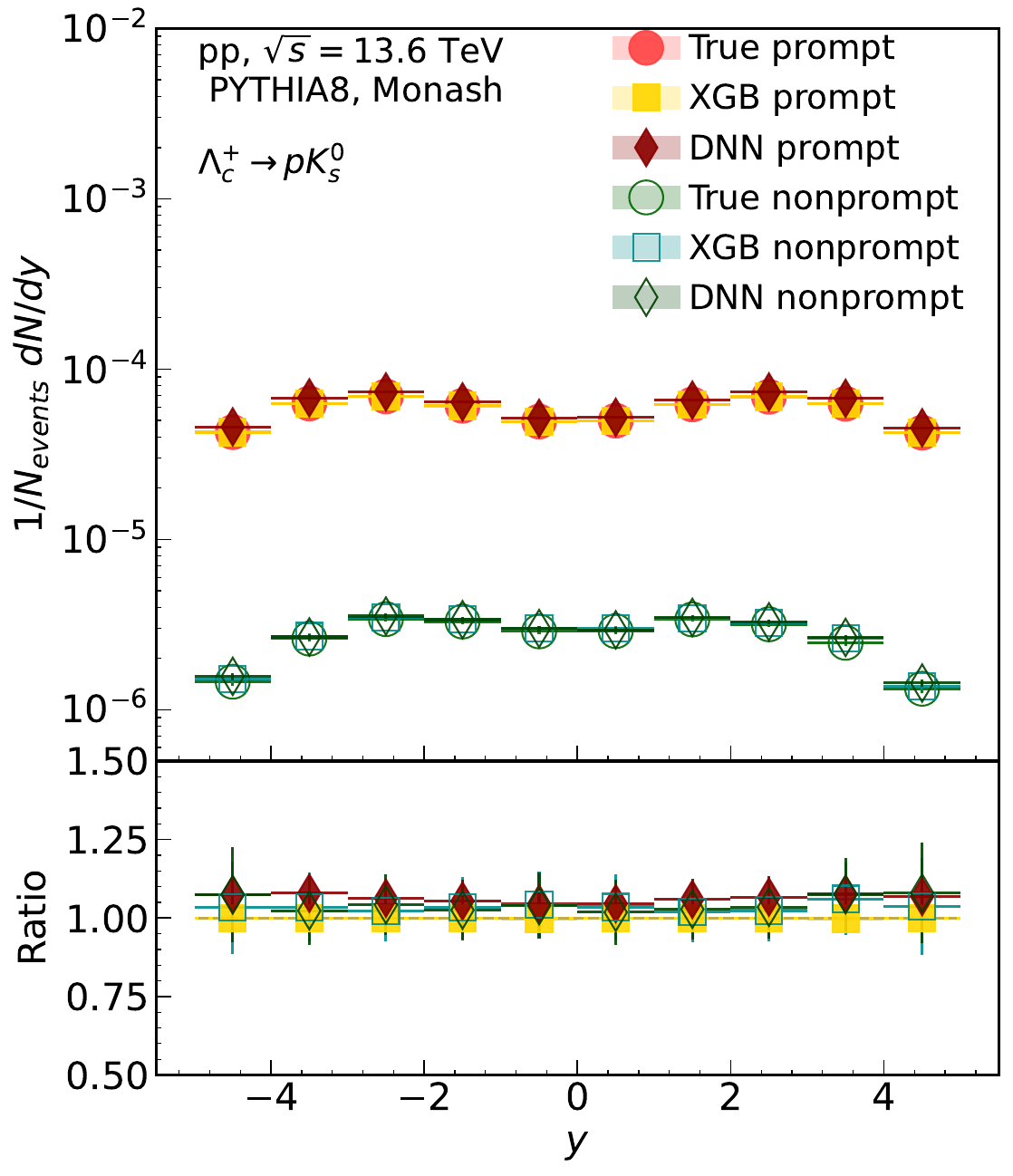}
    \caption{Rapidity distribution of $p_{\rm T}$-integrated prompt and nonprompt $\Lambda_c$-hadron in minimum bias $pp$ collisions at $\sqrt{s}~=~13.6$~TeV from the PYTHIA8 model along with the predictions from the XGB and DNN models. The bottom panel shows the ratio between predicted and true values for both XGB and DNN models.}
    \label{fig:yspectra}
\end{figure}

Figure~\ref{fig:yspectra} shows the $p_{\rm T}$-integrated rapidity distribution of prompt and nonprompt $\Lambda_c$-hadrons in minimum bias $pp$ collisions at $\sqrt{s}~=~13.6$~TeV from the PYTHIA8 model along with the predictions from the XGB and DNN models. The bottom panel shows the ratio between predicted and true values for both XGB and DNN models. Here, the yield is shown in a much wider rapidity coverage, \textit{i.e.,} $|y|<5.0$, to study the rapidity dependent behavior of prompt and nonprompt $\Lambda_c$-hadrons. As already described in the previous paragraph and shown in Fig.~\ref{fig:pTSpectra}, the nonprompt yield of $\Lambda_c^{+}$ is almost one order magnitude smaller than that of the prompt yield across all the rapidity bins. Both the prompt and nonprompt yields could be successfully reproduced by the XGB and DNN models, as evident from the bottom ratio plots, which fairly stay flat at unity. This suggests that the ML models can identify signal candidates and can tag them based on their topological production modes across a wide rapidity range without compromising the quality. However, the DNN model appears to slightly overpredict the yield in all the rapidity bins, while the XGB model is more accurate in recovering the true rapidity distribution.  

\begin{figure}[ht!]
    \centering
        \includegraphics[width=0.45\textwidth]{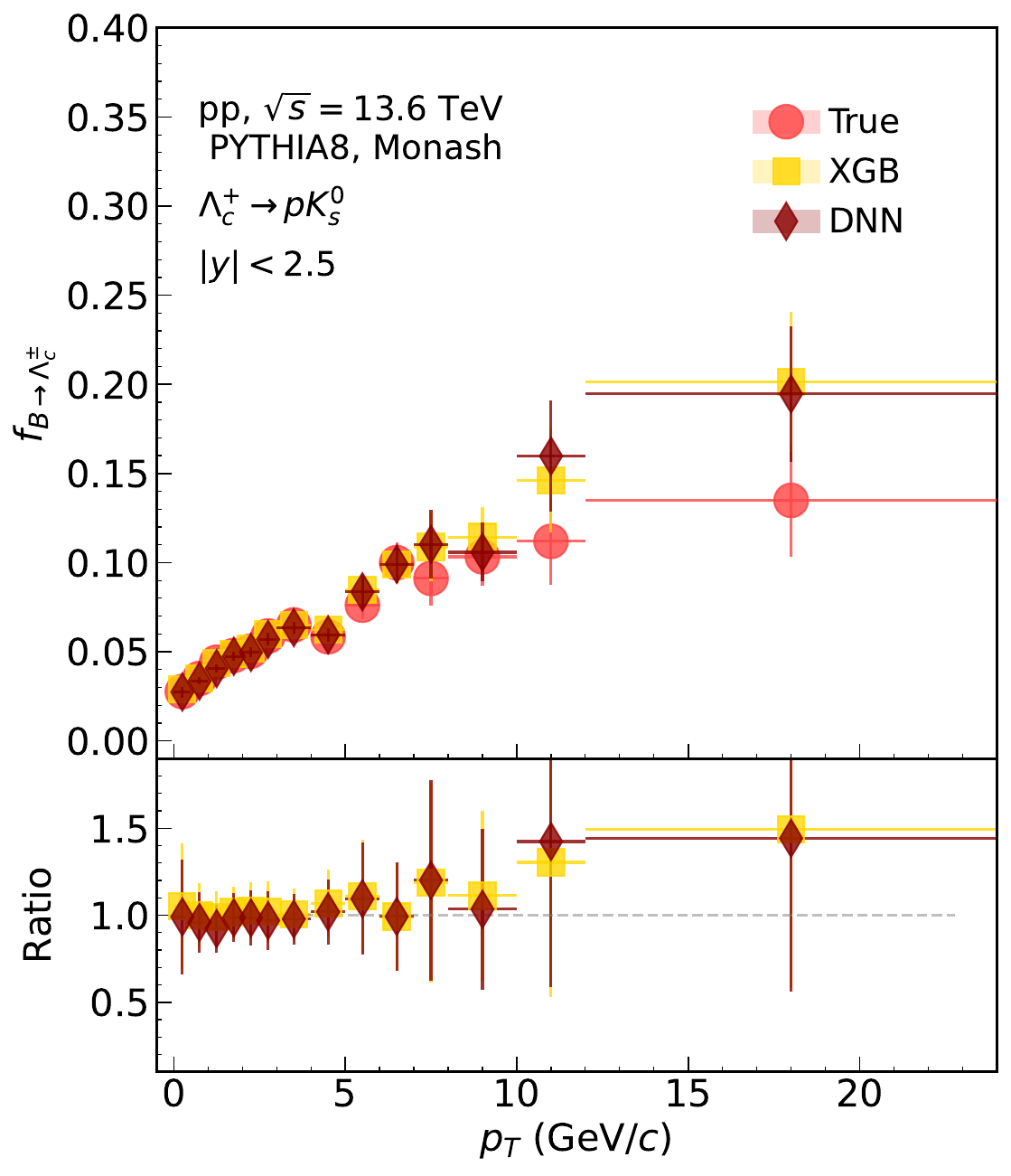}
    \caption{Transverse momentum dependence of nonprompt to inclusive $\Lambda_c^{+}$ ratio in minimum bias $pp$ collisions at $\sqrt{s}~=~13.6$~TeV measured in $|y|<2.5$. The true values from PYTHIA8, and the predictions from both XGB and DNN are shown. The bottom panel shows the ratio between predicted and true values for both XGB and DNN models.}
    \label{fig:fb_pt}
\end{figure}

\begin{figure}[ht!]
    \centering
        \includegraphics[width=0.45\textwidth]{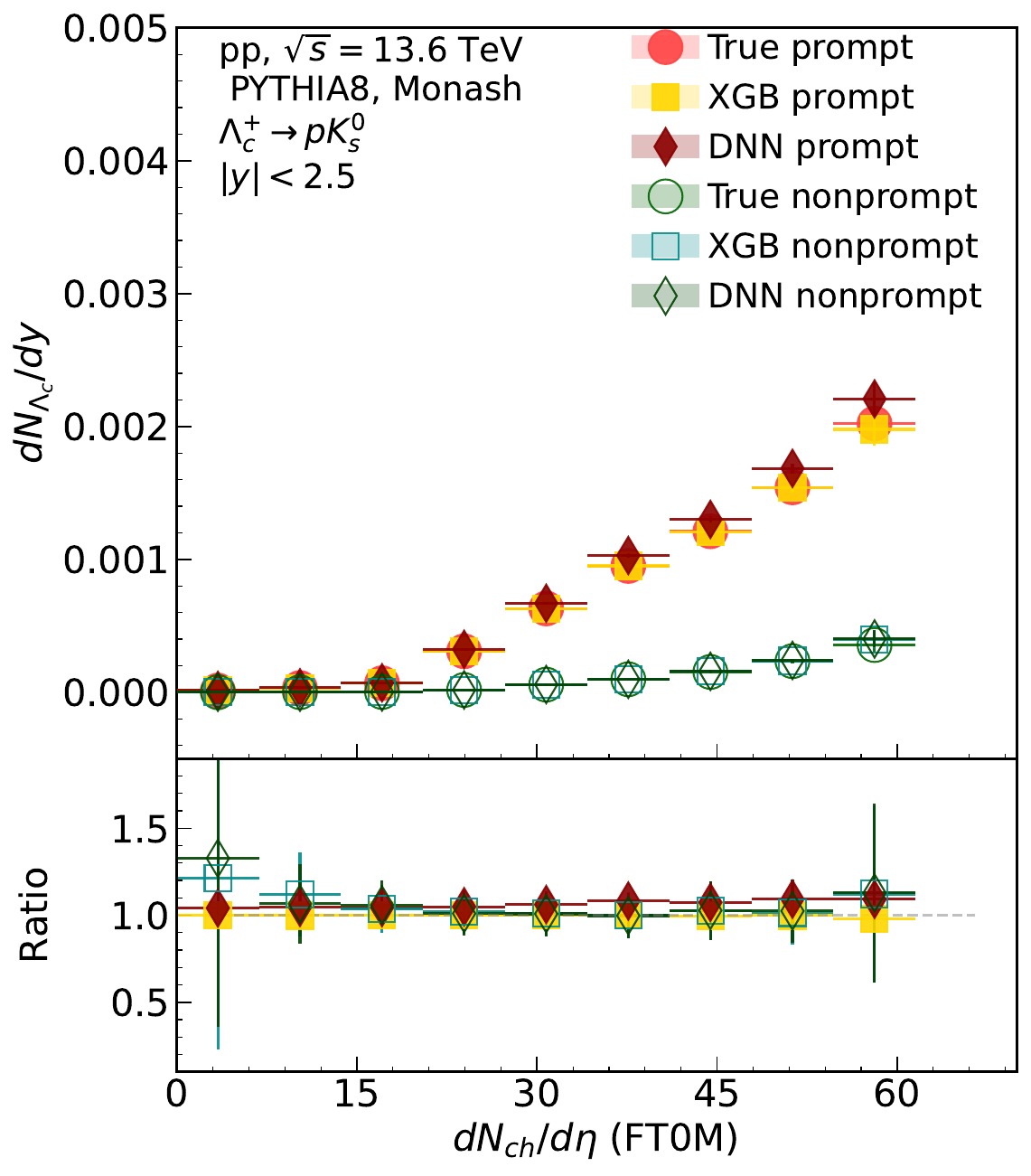}
    \caption{Multiplicity dependence of $p_{\rm T}$-integrated prompt and nonprompt $\Lambda_c^{+}$ yield in $pp$ collisions at $\sqrt{s}=13.6$~TeV in $|y|<2.5$. The true values from PYTHIA8, and the predictions from both XGB and DNN are shown. The estimation of charged-particle multiplicity density is performed in the ALICE-FT0M detector acceptance. The bottom panel shows the ratio between predicted and true values for both XGB and DNN models.}
    \label{fig:Lc_nCh}
\end{figure}

Figure~\ref{fig:fb_pt} shows the transverse momentum dependence of nonprompt to inclusive $\Lambda_c^{+}$ ratio in minimum bias $pp$ collisions at $\sqrt{s}~=~13.6$~TeV measured in $|y|<2.5$. The true values from PYTHIA8, and the predictions from both XGB and DNN are shown.  The nonprompt production fraction or $f_{B \rightarrow \Lambda_c^{\pm}}$ is used to calculate the nonprompt and prompt production cross sections ($\sigma_{\Lambda_c}$) from the inclusive cross section, as given below,
\begin{equation}
\begin{aligned}
    \sigma_{{\rm nonprompt}~\Lambda_{c}^{\pm}} &= f_{B \rightarrow \Lambda_c^{\pm}} \times \sigma_{{\rm incl.}~\Lambda_{c}^{\pm}}, \\
    \sigma_{{\rm prompt}~\Lambda_{c}^{\pm}} &= (1-f_{B \rightarrow \Lambda_c^{\pm}}) \times \sigma_{{\rm incl.}~\Lambda_{c}^{\pm}}.
\end{aligned}
\end{equation}
The value of $f_{B \rightarrow \Lambda_c^{\pm}}$ is found to be increasing linearly from low to intermediate-$p_{\rm T}$ up to $p_{\rm T}\simeq 6.0$~GeV/$c$. However, at high-$p_{\rm T}$, it seems to saturate and become independent of $p_{\rm T}$. The increase of nonprompt production fraction with increasing $p_{\rm T}$ can arise due to the fact that the production of charm quarks from the weak decay of beauty quarks is favored at high-$p_{\rm T}$ leading to an increase in the charm quark population.
This behavior is already observed in Fig.~\ref{fig:pTSpectra}, where the nonprompt $\Lambda_c^{+}$ yield approaches the prompt yield towards high-$p_{\rm T}$. Similar observations are also reported for other charm hadrons such as $D^0$ and $J/\psi$~\cite{Prasad:2023zdd, Goswami:2024xrx}. Now, by using machine learning, one can tag the $\Lambda_c$ directly from its decay candidates without any need of invariant mass fitting. Therefore, one can do an un-binned measurement of $f_{B \rightarrow \Lambda_c^{\pm}}$. Both XGB and DNN models are able to estimate the transverse momentum dependence of $f_{B \rightarrow \Lambda_c^{\pm}}$. The ratios between the predicted values from the models and the true values are almost equal to unity, which indicates the accuracy of the models.

Finally, the $p_{\rm T}$-integrated prompt and nonprompt $\Lambda_c^{+}$ yield as a function of charge-particle multiplicity density in $pp$ collisions at $\sqrt{s}=13.6$~TeV is presented in Fig.~\ref{fig:Lc_nCh}. The estimation of charged-particle multiplicity density is performed in the ALICE-FT0M detector acceptance, which covers the forward pseudorapidity regions in $3.5<\eta<4.9$ and $-3.3< \eta < -2.1$. The estimation of the $\Lambda_c^{+}$ yield is performed in $|y|<2.5$, to avoid double counting of tracks and autocorrelation. For $dN_{\rm ch}/d\eta \leq 18$, both the yield of prompt and nonprompt $\Lambda_c^{+}$ are fairly comparable. However, the yield of prompt $\Lambda_c^{+}$ begins to increase sharply for $dN_{\rm ch}/d\eta > 18$ and keeps increasing almost linearly towards the high multiplicity events. However, the yield of nonprompt $\Lambda_c^{+}$ has a very mild increase towards the high multiplicity events. At the highest charged-particle multiplicity bin ($dN_{\rm ch}/d\eta \simeq 57$), the prompt $\Lambda_c^{+}$ yield is approximately seven times higher than the nonprompt yield. Both XGB and DNN models successfully estimate the $p_{\rm T}$-integrated prompt and nonprompt $\Lambda_c^{+}$ yield as a function of charge-particle multiplicity density as seen from the bottom ratio plot, which shows excellent accuracy between the true and predicted values.

\section{Summary}
\label{sec:summary}
This study details the use of XGBoost and Deep Neural Network based multi-class classification models to perform an unbinned track-level identification and tagging of prompt and nonprompt $\Lambda_c$-hadrons from its three body final state decay. The machine learning models are trained with $pp$ collisions at $\sqrt{s} = 13.6$~TeV data simulated with PYTHIA8 (Monash) model. To mimic the experimental environment, these supervised classification models are trained on a few experimentally available features of the decay candidates such as invariant mass, pseudoproper decay length, the distance of closest approach of $\Lambda_c^+$ and the intermediate $K^0_s$. Both the models achieve almost $100\%$ accuracy in background rejection and $99\%$ accuracy in tagging the prompt candidates of $\Lambda_c$-hadron. Therefore, the identified signal candidates are free from the combinatorial background, therefore, can be used in several analyses by reducing the steps involving the invariant mass reconstruction. This is specifically useful for small collision systems where the S/B is low for charm-hadrons such as $\Lambda_c^+$.
Additionally, the XGBoost model outperforms the DNN model in tagging the nonprompt $\Lambda_c^+$ candidates with $96\%$ accuracy, whereas the DNN model could achieve a $94\%$ accuracy.

\section*{Acknowledgements}
We would like to thank the Jyväskylä ALICE group for fruitful discussions.
We acknowledge CSC—IT Center for Science in Espoo, Finland, for the allocation of computational resources. NM and DJK are supported by the Academy of Finland through the Center of Excellence in Quark Matter (Grant No. 346328), and HH through Grant No. 346327.

\end{document}